\documentclass[10pt,twocolumn]{article}
\usepackage[a4paper,margin=1in]{geometry}
\usepackage{times}
\usepackage{hyperref}
\usepackage{graphicx}
\usepackage{amsmath}
\usepackage{amssymb}
\usepackage{cite}
\usepackage{booktabs}
\usepackage{lmodern}
\usepackage{color}
\usepackage{inputenc}

\title{\bf AuthorMist: Evading AI Text Detectors with Reinforcement Learning}
\date{}

\author{
    \begin{tabular}{c}
        \textbf{Isaac David} ~ 
        \textbf{Arthur Gervais} \\
        University College London
    \end{tabular}
}

\begin{document}
\maketitle

\begin{abstract}
    In the age of powerful AI-generated text, automatic detectors have emerged to identify machine-written content.
    This poses a threat to author privacy and freedom, as text authored with AI assistance may be unfairly flagged.
    We propose \textit{AuthorMist}, a novel reinforcement learning-based system to transform AI-generated text into human-like writing.
    AuthorMist leverages a 3-billion-parameter language model as a backbone, fine-tuned with Group Relative Policy Optimization (GPRO) to paraphrase text in a way that evades AI detectors.

    Our framework establishes a generic approach where external detector APIs (GPTZero, WinstonAI, Originality.ai, etc.) serve as reward functions within the reinforcement learning loop, enabling the model to systematically learn outputs that these detectors are less likely to classify as AI-generated.
    This API-as-reward methodology can be applied broadly to optimize text against any detector with an accessible interface.
    Experiments on multiple datasets and detectors demonstrate that AuthorMist effectively reduces the detectability of AI-generated text while preserving the original meaning. Our evaluation shows attack success rates ranging from 78.6\% to 96.2\% against individual detectors, significantly outperforming baseline paraphrasing methods. AuthorMist maintains high semantic similarity (above 0.94) with the original text while successfully evading detection. These results highlight limitations in current AI text detection technologies and raise questions about the sustainability of the detection-evasion arms race.
\end{abstract}
 
\section{Introduction}
The rise of large language models (LLMs) capable of producing human-like text has led to increased use of AI assistants in writing.
Students, content creators, and professionals now often rely on AI-generated drafts for efficiency.
However, this has spurred the development of AI text detectors -- tools that attempt to distinguish AI-written text from human-written text.
Systems like GPTZero and WinstonAI claim high accuracy in flagging machine-generated content \cite{Tian2023GPTZero,Winston2023Detector}.
Their widespread use in education and publishing poses a challenge to authorship privacy: an author who uses AI assistance might be unfairly scrutinized or penalized if their text is marked as ``AI-generated.''
This encroaches on author freedom and could discourage beneficial uses of AI in writing.

Recent studies have shown that current detectors are far from foolproof.
Detectors often rely on statistical signatures of AI text (e.g., likelihood under a language model or the presence of out-of-distribution tokens).
Yet, even simple transformations to the text can significantly degrade detector performance.
For example, adding extraneous spaces, introducing misspellings, or replacing characters with look-alike symbols can drop detection accuracy by over 30\%.
Paraphrasing the content in different words can often render AI-generated text ``undetectable'' to many models \cite{Perkins2024Evasion}.
This cat-and-mouse dynamic highlights a need for robust methods to \emph{evade} detection, both to test detector reliability and to empower users to protect their privacy.

In this paper, we introduce \textbf{AuthorMist}, a reinforcement learning (RL) approach to automatically disguise AI-generated text as human-written.
Rather than relying on crude text perturbations, AuthorMist learns a policy to paraphrase text into a more human-like style while preserving its meaning.
We use RL with external AI detectors in a loop: the model is rewarded when the output text is rated as human-written by detectors.
Over time, the model learns to produce texts that systematically fool detectors into low ``AI likelihood'' scores.

Our contributions are as follows:
\begin{itemize}
    \item We formulate the evasion of AI text detectors as a reinforcement learning problem. We pioneer a novel ``API-as-reward'' methodology, being the first to integrate external detector APIs directly into the reinforcement learning loop with Group Relative Policy Optimization (GRPO) \cite{Shao2024GRPO} for NLP tasks. 
    \item Our system, AuthorMist, is designed and built on a 3B-parameter Transformer model fine-tuned with this innovative approach.
    \item We thoroughly evaluate AuthorMist on benchmark datasets of AI-generated text. We test against popular blackbox detectors (GPTZero, WinstonAI, Originality.ai, Sapling) and popular open-source models (HelloSimpleAI \cite{guo-etal-2023-hc3} and OpenAI's Roberta Detector \cite{solaiman2019release}). Results show AuthorMist dramatically reduces detection rates compared to baseline AI text, outperforming conventional paraphrasers.
\end{itemize}

\section{Background}
\subsection{AI Text Detectors and Adversarial Evasion}
The concern over machine-generated misinformation and academic dishonesty has led to a proliferation of \textit{AI text detectors}.
These tools use various strategies to identify telltale signs of AI authorship.
Some detectors are neural classifiers trained on human vs.\ AI text examples (e.g., fine-tuned RoBERTa models), while others use statistical metrics like entropy or perplexity (e.g., GLTR \cite{Gehrmann2019GLTR} and DetectGPT \cite{Mitchell2023DetectGPT}).
Commercial services such as GPTZero \cite{Tian2023GPTZero} and WinstonAI \cite{Winston2023Detector} combine multiple features and claim high accuracy and precision.
For instance, GPTZero has been noted for its robustness to character-level attacks, and WinstonAI advertises $>$99\% accuracy on certain benchmarks.

Perplexity, in particular, has emerged as a key statistical signature that differentiates AI-generated from human-written text. AI-generated content typically exhibits lower perplexity (higher predictability) than human writing, as language models tend to produce more statistically likely token sequences. This predictability pattern is a common detection vector that many AI text detectors exploit. Human writing, in contrast, often contains more surprising word choices and unusual constructions, resulting in higher perplexity scores.

Despite these claims, detectors are inherently engaged in an arms race with generation methods.
Researchers have demonstrated that even straightforward \textit{adversarial attacks} can significantly lower detector confidence.
Creo \textit{et al.} (2024)~\cite{creo2025silver} showed that inserting innocuous Unicode homoglyphs or random punctuation can break detectors, causing their performance to degrade sharply.
In academic contexts, Perkins \textit{et al.} \cite{Perkins2024Evasion} evaluated six major detectors and found their average accuracy (already a modest 39.5\%) dropped to 17.4\% when faced with texts lightly modified to evade detection.
A study from Google Research (Krishna \textit{et al.}, 2023)~\cite{dipper} introduced a paraphrasing model called Dipper that rewrote AI text and was able to bypass many detectors nearly universally.
Dipper demonstrated impressive evasion capabilities, reducing DetectGPT's detection accuracy from 70.3\% to just 4.6\% at a constant false positive rate of 1\%. However, Dipper relies on a one-step paraphrasing approach rather than an iterative optimization process, potentially limiting its ability to adapt to evolving detection methods.
These so-called ``AI humanizer'' tools have proliferated online~\cite{Masrour2025DAMAGE}, offering students and writers the means to automatically transform AI outputs to evade detection.

However, naive adversarial tricks often come at the cost of readability or fidelity.
Text with missing articles or odd spellings might evade an algorithm yet appear obviously flawed to a human reader.
What is needed is a method to \textit{paraphrase} AI content into a form that both preserves its meaning and appears human-crafted and natural in style.
This is a non-trivial task: it requires altering the subtle statistical signature of the text without introducing errors or unnatural phrasing. Semantic preservation is particularly challenging, as even small changes in wording can significantly alter meaning, especially in technical or specialized domains.

\subsection{Reinforcement Learning in Text Generation}
Reinforcement learning has emerged as a powerful paradigm to fine-tune language models for objectives that are difficult to express with direct supervision.
RL from human feedback (RLHF) has been used to align LLMs with human preferences, using methods like Proximal Policy Optimization (PPO)~\cite{schulman2017proximal} to adjust a model based on reward signals from human evaluators or learned reward models.
In our scenario, the ``preference'' is that text should appear human-written, and detectors play the role of automated evaluators of this property.

A recent line of work applies pure RL to language model training without any supervised fine-tuning.
Notably, DeepSeek's R1-Zero model was trained via large-scale reinforcement learning alone, aiming to improve the model's reasoning ability \emph{de novo} \cite{DeepSeek2025R1}.
DeepSeek-R1 employed a custom algorithm called Group Relative Policy Optimization (GRPO) to stabilize training.
GRPO is a variant of PPO that forgoes a separate value critic; instead, it evaluates a group of sampled outputs to compute a baseline reward, using their average as a reference.
This relative reward approach reduces variance and resource usage by eliminating the need to train a value network \cite{Shao2024GRPO}.
GRPO has been shown to be effective at improving certain capabilities of LLMs (e.g., mathematical reasoning in DeepSeekMath \cite{Shao2024GRPO}) with less overhead than PPO.
However, purely RL-trained models can sometimes exhibit peculiarities: DeepSeek's R1-Zero, e.g., achieved strong reasoning performance but at the expense of fluent language, occasionally producing mixed-language outputs and lower readability.

To maintain linguistic quality during RL fine-tuning, many approaches incorporate a Kullback-Leibler (KL) divergence~\cite{kullback1951information} penalty that prevents the policy from deviating from the initial language model. This regularization technique is crucial for text generation tasks, as it helps preserve fluency and coherence while allowing the model to optimize for task-specific objectives. In the context of evading AI text detection, KL divergence regularization helps ensure that the paraphrased text remains natural and readable while becoming less detectable.

While RL offers powerful optimization capabilities, alternative approaches like Supervised Fine-Tuning (SFT)~\cite{gunel2020supervised} are also commonly used for text transformation tasks. SFT involves training a model on paired examples of input and desired output text, which can be effective for straightforward paraphrasing. However, SFT typically lacks the ability to directly optimize for complex objectives like detector evasion, as it cannot incorporate non-differentiable reward signals from external systems. This limitation makes RL particularly well-suited for our task of evading AI text detectors.

These insights inform our approach.
In this work, we adopt an RL strategy (in contrast to simple one-step paraphrasing) to iteratively adjust a generative model towards a hard-to-define objective: evading detectors.
By leveraging GRPO in our training loop, we maintain efficiency and stable convergence.

\section{Design}

In the following, we describe the design of AuthorMist in detail.

\subsection{Reward Modeling}
The AuthorMist system requires feedback from AI text detectors to guide its RL process. Our design incorporates multiple detectors to provide robust and varied feedback within the RL loop. With also train a variety of AuthorMist models to evade different detectors, we ensure that our system learns to evade a diverse range of detection algorithms, rather than overfitting to a single detector's weaknesses. For example, we train AuthorMist against originality.ai as a dedicated version. By integrating both commercial and open-source detectors, we ensure that our system learns to evade a diverse range of detection algorithms, rather than overfitting to a single detector's weaknesses.

The detector selection strategy is guided by several key design principles:

\paragraph{Diversity of Detection Approaches}
We deliberately incorporate detectors that employ different underlying techniques for identifying AI-generated text. This includes:
\begin{itemize}
    \item Statistical detectors that analyze token distributions and perplexity patterns
    \item Neural classifiers trained on large datasets of human vs. AI text
    \item Hybrid systems that combine multiple detection signals
\end{itemize}
This diversity ensures that AuthorMist learns to address multiple detection vectors simultaneously rather than exploiting a single weakness.

\paragraph{Representativeness}
The selected detectors should represent those commonly used in real-world scenarios, such as academic integrity checks, content moderation systems, and publishing workflows. This ensures that AuthorMist's evasion capabilities are practically useful in contexts where users might face AI detection.

\paragraph{API Stability and Reliability}
For effective training, detectors must provide consistent and reliable feedback. Our design accounts for API rate limits, response time variability, and potential changes in detector behavior over time. The system includes mechanisms to handle temporary API failures and ensure training stability despite these challenges.

\paragraph{Feedback Granularity}
All detectors provide continuous probability scores over simple binary classifications. Continuous scores offer richer training signals that allow the RL algorithm to make incremental improvements, whereas binary feedback provides less guidance for optimization.

\subsection{System Architecture}
\begin{figure*}[t]
    \centering
    \includegraphics[width=1.02\textwidth]{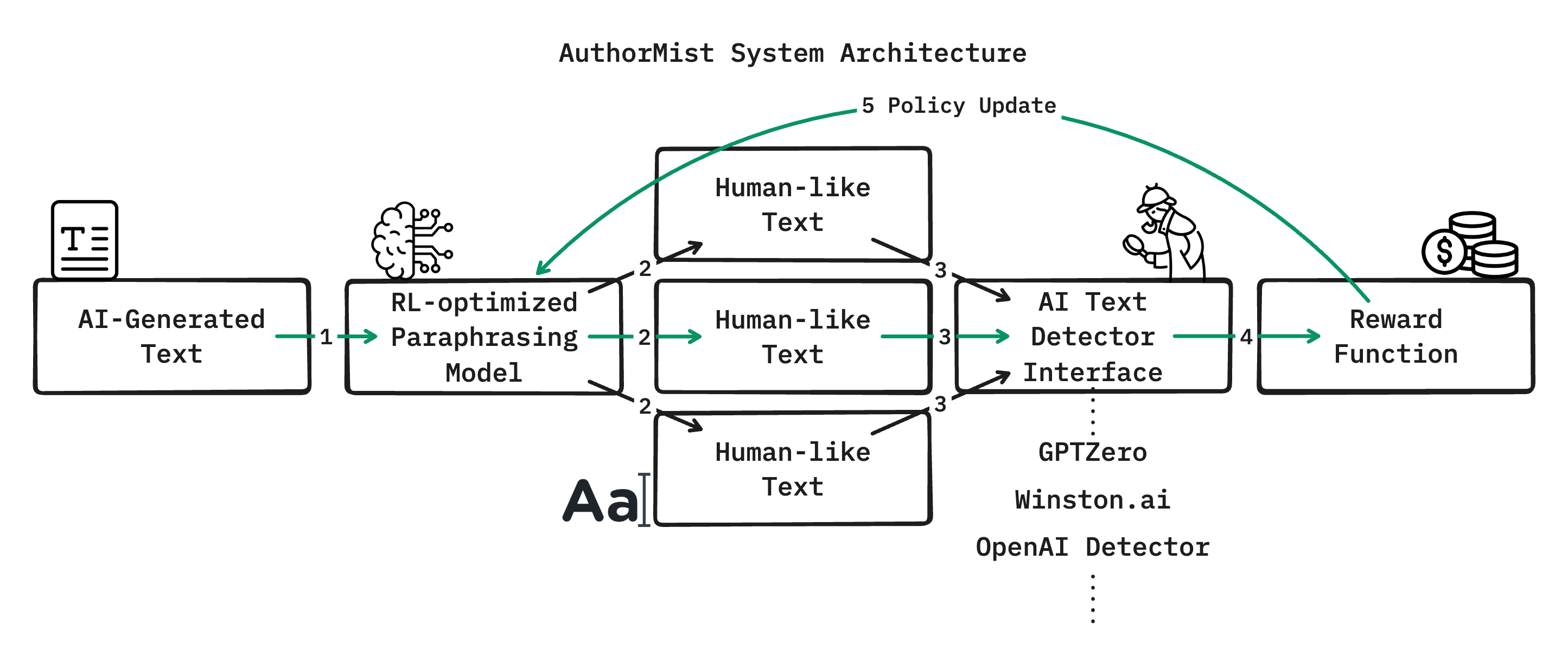}
    \caption{AuthorMist system architecture. The system takes AI-generated text as input and processes it through an RL-optimized paraphrasing model trained to minimize detector scores, producing human-like text that preserves the original meaning while evading detection.}
    \label{fig:system_architecture}
\end{figure*}

AuthorMist consists of two main components: (1) a base language model that serves as the paraphrasing policy, and (2) a reinforcement learning framework that optimizes this policy using detector feedback. Figure \ref{fig:system_architecture} illustrates the complete system architecture and workflow.

During inference, AI-generated text is fed into the RL-trained paraphrasing model, which transforms it to minimize detection probability. The architecture consists of several key components working in concert:

\paragraph{Base Language Model}
The foundation of AuthorMist is a pre-trained language model that provides the initial paraphrasing capabilities. We select Qwen2.5-3B Instruct~\cite{qwen2.5} as the base model due to its strong balance between performance and efficiency. This model is selected for its ability to generate fluent and coherent text while being computationally efficient enough to allow for extensive reinforcement learning. The relatively compact size of Qwen2.5-3B enables faster training iterations while still maintaining high-quality text generation capabilities. The base model serves as the initial policy that will be optimized through RL.

\paragraph{Detector Interface Layer}
This component provides a standardized interface to multiple AI text detectors. It handles the communication with external detector APIs, manages rate limiting, and normalizes the various detector outputs (probability scores or binary classifications) into a consistent format that can be used by the reward function. This abstraction layer allows AuthorMist to easily incorporate new detectors as they become available.

\paragraph{RL Training Loop}
The core of AuthorMist is its reinforcement learning loop, which iteratively improves the paraphrasing policy. For each training example, the system:
\begin{enumerate}
    \item Takes an AI-generated text as input
    \item Generates multiple paraphrased versions using the current policy
    \item Submits these versions to the detector interface
    \item Computes rewards based on detector feedback
    \item Updates the policy parameters to maximize expected rewards
\end{enumerate}

\subsection{Reward Function Definition}
The reward function is designed to quantitatively measure the success of AuthorMist in evading AI-generated text detection. Given a set of detectors \(D = \{d_1, d_2, \dots, d_k\}\), each detector \(d_j\) outputs either a probability score \(P_{d_j}(Y)\) in the range \([0, 1]\), indicating the likelihood of text \(Y\) being AI-generated, or a binary classification.

When detectors output a continuous probability score, the reward \(R(X,Y)\) for transforming an AI-generated input \(X\) into output \(Y\) is computed as follows:
\[
    R(X,Y) = 1 - \frac{1}{k}\sum_{j=1}^{k} P_{d_j}(Y)
\]

Thus, the model receives a higher reward when outputs are classified as more human-like (lower probability of being AI-generated).

\subsection{Training Data Design}
The design of training data is a critical component of AuthorMist's effectiveness. Our approach to training data creation and selection is guided by several principles: diversity of AI-generated sources (incorporating text from multiple state-of-the-art language models to learn generalizable paraphrasing strategies), comprehensive domain coverage (spanning academic, creative, technical, and conversational styles to ensure effectiveness across contexts), varied length distribution (including short paragraphs to multi-paragraph passages to address detection algorithms' varying behaviors based on text length), and the inclusion of human reference texts (which serve as implicit examples of the target distribution, influencing the initial policy and providing a stylistic anchor for KL divergence regularization without being directly used in the reward function).

For our training dataset, we selected 10,000 human-written abstracts from the CheckGPT dataset~\cite{liu2024detectability} and created corresponding AI-generated versions by paraphrasing them using state-of-the-art language models including GPT-4o, Claude 3.5 Sonnet, and Gemini 1.5. This approach allowed us to create human:AI text pairs where we have precise control over the AI generation process. The dataset spans multiple domains including Computer Science (CS), Humanities and Social Sciences (HSS), and Physics (PHX), ensuring that our model learns to paraphrase effectively across different subject areas and writing styles.

The training samples vary in length from 100 to 500 words, with a median length of approximately 250 words. This length distribution is particularly important as detection algorithms often exhibit different behaviors based on text length, with some detectors performing better on longer passages where statistical patterns become more apparent. By training on a diverse range of text lengths, AuthorMist learns to adapt its paraphrasing strategies appropriately for both short and long content.

\subsection{RL with GRPO}
To optimize AuthorMist's rewriting capabilities, we employ Group Relative Policy Optimization (GRPO). During training, for each input \(X_i\), we sample multiple paraphrased outputs \(\{Y_{i1}, Y_{i2}, \dots, Y_{iG}\}\) and compute their corresponding rewards. GRPO calculates the baseline reward \(b_i\) as:
\[
    b_i = \frac{1}{G}\sum_{j=1}^{G} R(X_i, Y_{ij})
\]

The advantage \(A_{ij}\) for each sample is determined by:
\[
    A_{ij} = R(X_i, Y_{ij}) - b_i
\]

The model parameters \(\theta\) are then updated by maximizing the following objective function:
\[
    J(\theta) \approx \frac{1}{N}\sum_{i=1}^{N}\frac{1}{G}\sum_{j=1}^{G} A_{ij}\sum_{t}\log \pi_\theta(y_{ij,t}\mid X_i, y_{ij,<t})
\]

To maintain linguistic fluency and prevent unnatural text artifacts, we incorporate a Kullback-Leibler (KL) divergence penalty in our optimization objective, keeping the updated policy distribution close to the initial policy.

\paragraph{GRPO Design Considerations}
We selected GRPO over other RL algorithms like PPO and DPO for its computational efficiency (eliminating separate value networks to reduce overhead), stability in high-variance text generation (using relative reward mechanisms for more stable gradients), sample efficiency (learning from multiple paraphrased versions simultaneously to accelerate convergence), and adaptability to evolving detectors (employing relative comparison approaches that remain robust as detection methods change).

\paragraph{KL Divergence Regularization}
A critical design element is the KL divergence penalty that prevents the policy from deviating too far from the initial language model, serving multiple purposes:

\begin{itemize}
    \item Preserving linguistic quality by anchoring to a well-trained language model
    \item Preventing degenerate solutions that might evade detectors but produce unnatural text
    \item Ensuring semantic preservation by limiting how much the model can alter the input
\end{itemize}

The strength of this regularization is carefully balanced to allow sufficient freedom for the model to learn effective evasion strategies while maintaining text quality.

\section{Evaluation}
In this section, we present the results of our evaluation of AuthorMist.

\subsection{Dataset and Metrics}

\paragraph{Dataset:}
For our evaluation, we randomly sampled $300$ entries from the Xsum dataset~\cite{narayan2018don} (A dataset of human-authored news article summaries), comprising a diverse range of topics and writing styles. These human-authored summaries were paraphrased using popular state-of-the-art LLMs (GPT-4o, Claude 3.5 Sonnet, and Gemini 1.5), creating a balanced dataset of $300$ human-written and AI-generated text pairs for evaluation. The human-written texts in these pairs serve as our control group, establishing natural baseline detection rates for each detector. %These baseline measurements are crucial for two purposes: (1) calculating false positive rates to ensure detector calibration, and (2) providing reference points for semantic similarity and fluency metrics when evaluating how well our paraphrasing models preserve the original meaning and readability while evading detection.

\paragraph{Paraphrasing Models}
For our evaluation, we developed six distinct paraphrasing models using GRPO, with each model specifically trained to evade one particular AI text detector. For example, one model was optimized to bypass GPTZero, another for OpenAI's detector, and so on. We then used these specialized models to paraphrase our collection of AI-generated texts. To thoroughly assess how well our approach generalizes, we didn't limit testing to just the detector each model was trained against. Instead, we evaluated each model's paraphrased outputs against all six detectors in our test suite. This comprehensive cross-evaluation allowed us to determine whether a model trained to evade one detector could successfully transfer those capabilities to other detection systems. Throughout this process, we carefully recorded the probability scores (indicating likelihood of AI generation) returned by each detector for subsequent analysis.

\paragraph{Evaluation Metrics}
To quantitatively measure the effectiveness of GRPO-trained models in evading AI detection, we employed the following key metrics:
\begin{itemize}
    \item \textbf{Attack Success Rate (ASR)}: The percentage of AI-generated texts misclassified as human after paraphrasing. Higher ASR indicates better evasion capability.
    \item \textbf{Area Under the Receiver Operating Characteristic Curve (AUROC)}: A threshold-independent metric that quantifies a detector's discrimination ability by plotting true positive rate against false positive rate across all possible classification thresholds~\cite{fawcett2006introduction}.
          AUROC ranges from $0$ to $1$, where $0.5$ represents random guessing. In our context, lower AUROC values (closer to $0.5$) indicate that detectors struggle to distinguish between human and AI-generated text, confirming successful evasion.
    \item \textbf{F1-score}: The F1-score is the harmonic mean of precision and recall. Lower F1-scores indicate better evasion performance.
\end{itemize}

To quantitatively evaluate semantic preservation between original and paraphrased texts, we employed the following semantic similarity metric:
\begin{itemize}
    \item \textbf{Semantic Similarity}: We utilized the E5 small embedding model~\cite{wang2022text} to compute cosine similarity between the original and paraphrased text embeddings. This state-of-the-art text embedding model was specifically designed for semantic similarity tasks through weakly-supervised contrastive pre-training, providing a reliable measure of how well our paraphrasing preserves the original meaning. Higher similarity scores (closer to 1.0) indicate better semantic preservation.
\end{itemize}

To evaluate the fluency of the paraphrased texts, we used the following metrics:
\begin{itemize}
    \item \textbf{Perplexity}: We compute the perplexity of the paraphrased texts with GPT2 as done by related works~\cite{dipper}. Lower perplexity indicates better fluency, as it measures how well a language model can predict the text. Fluent, natural text results in lower perplexity scores.
\end{itemize}

\subsection{Implementation Details}

\paragraph{Detector Selection}
For our implementation, we selected four commercially available AI-generated text detectors—GPTZero, Winston.ai, Sapling, and Originality.ai—due to their advertised high performance and widespread use in academic and professional settings. Additionally, we incorporated two popular open-source detectors from HuggingFace: HelloSimpleAI and OpenAI RoBERTa-based detectors. The open-source models were selected based on their monthly download statistics, reflecting their current popularity and community trust.

\paragraph{Base Model Selection}
For our experiments, we selected the Qwen2.5-3B Instruct model as the foundational language model. The choice of this model is driven by its strong balance between computational efficiency and language proficiency, making it suitable for rapid iterative experiments and fine-tuning via reinforcement learning.

\paragraph{Training Configuration}
For our GRPO implementation, we set the learning rate to \(5\times10^{-5}\) and applied a KL divergence penalty with \(\beta=0.001\). The model was trained for a single epoch (714 Steps) using eight NVIDIA H100 GPUs, with training completed in two hours, totaling 16 GPU-hours. In total, we trained six separate models, each using a single detector as the reward function.

\paragraph{Inference Setup}
AuthorMist processes input text in chunks of up to 512 tokens to manage memory constraints and ensure consistent quality across longer documents. For each chunk, the system generates $8$ candidate paraphrases and selects the one with the lowest average detector score.

\subsection{Results}
In the following, we present the results for each of the six GRPO-trained models, comparing their performance across all detectors and metrics.

\paragraph{Attack Success Rate (ASR)}
Table~\ref{tab:cross_detector_asr} presents the Attack Success Rate (ASR) for each model-detector combination. The ASR metric quantifies evasion effectiveness by measuring the percentage of AI-generated texts successfully misclassified as human-written after paraphrasing. The baseline column shows the detection performance on unmodified AI-generated text, demonstrating the significant detection challenges faced by unaltered AI-generated content.

\begin{table*}[htbp]
    \centering
    \caption{Cross-Detector Attack Success Rate (ASR \%) for AuthorMist Bypasser Models. Columns represent the detector used as a reward function during training (AuthorMist trained against), while rows represent evaluated detectors. Diagonal values indicate evaluations on paraphrased texts specifically targeting the same detector. Higher ASR indicates better success in bypassing the detector.}
    \label{tab:cross_detector_asr}
    \resizebox{\textwidth}{!}{%
        \begin{tabular}{lccccccc}
            \toprule
                                   & \multicolumn{7}{c}{\textbf{AuthorMist Trained Against}}                                                                                                                             \\
            \cmidrule(lr){2-8}
            \textbf{Eval Detector} & \textbf{Baseline}                                       & \textbf{GPTZero} & \textbf{OpenAI} & \textbf{Hello SimpleAI} & \textbf{Sapling} & \textbf{Originality} & \textbf{Winston} \\
            \midrule
            GPTZero                & 4.00                                                    & \textbf{92.33}   & 2.00            & 10.00                   & 41.33            & 83.67                & 86.00            \\
            OpenAI                 & 100.00                                                  & 19.33            & \textbf{95.67}  & 42.33                   & 85.00            & 93.33                & 52.33            \\
            Hello SimpleAI         & 49.67                                                   & 82.33            & 81.67           & \textbf{97.67}          & 94.67            & 100.00               & 98.33            \\
            Sapling                & 2.00                                                    & 99.00            & 2.33            & 23.00                   & \textbf{98.33}   & 100.00               & 97.00            \\
            Originality            & 0.00                                                    & 44.00            & 0.00            & 0.33                    & 8.33             & \textbf{94.00}       & 68.33            \\
            Winston                & 33.33                                                   & 100.00           & 12.33           & 72.67                   & 99.00            & 100.00               & \textbf{100.00}  \\
            \midrule
            \textbf{Mean ASR}      & --                                                      & 72.83            & 32.33           & 41.00                   & 71.11            & 95.17                & 83.67            \\
            \bottomrule
        \end{tabular}}
\end{table*}

Our results reveal the substantial effectiveness of our GRPO-trained models. For example, the model trained against GPTZero achieved a $92.33\%$ success rate against its target detector, indicating that more than $92\%$ of AI-generated texts were misclassified as human after paraphrasing. Comparable high-performance results appear across the diagonal of the table, with all models achieving at least $94\%$ ASR against their target detectors, with the exception of the GPTZero-trained model.

The Originality.ai-trained model exhibits the strongest overall performance, with a mean ASR of $95.17\%$ across all detectors, suggesting this model developed generalizable evasion strategies applicable to multiple detection systems. The Winston.ai-trained model also demonstrates robust cross-detector performance ($83.67\%$ mean ASR). It is worth noting that several models achieve a perfect $100\%$ ASR against certain detectors, effectively rendering these detection systems completely ineffective against our paraphrasing approach.

\paragraph{Area Under the Receiver Operating Characteristic Curve (AUROC)}
Table~\ref{tab:cross_detector_auroc} presents the Area Under the Receiver Operating Characteristic Curve (AUROC) for each model-detector combination. The AUROC metric provides a comprehensive assessment of detector performance across all possible classification thresholds, with values closer to 0.5 indicating that detectors struggle to distinguish between human and AI-generated text (successful evasion).

The baseline column demonstrates the strong detection capabilities of unmodified AI text detectors, with most achieving AUROC values above 0.95, indicating high discrimination ability. In contrast, our GRPO-trained AuthorMist models significantly degrade detector performance. For example, the Hello SimpleAI-trained model achieved an AUROC of 0.086 against its target detector, effectively reducing the detector to worse-than-random performance. Similarly impressive results appear across the diagonal, with all models achieving substantial AUROC reductions against their target detectors.

Cross-detector performance reveals interesting patterns of generalization. The AuthorMist Originality model (trained against Originality.ai) demonstrates remarkable transfer capabilities, achieving an AUROC of 0.070 against HelloSimpleAI despite not being explicitly trained on it. The AuthorMist Winston model (trained against Winston.ai) also shows strong generalization, with a median AUROC of 0.58 across all detectors. These results suggest that certain evasion strategies learned against one detector can transfer to others, highlighting potential similarities in underlying detection mechanisms across different systems.

% \paragraph{AUROC Distribution}
\begin{figure*}
    \centering
    \begin{tabular}{cc}
        \includegraphics[width=0.45\textwidth]{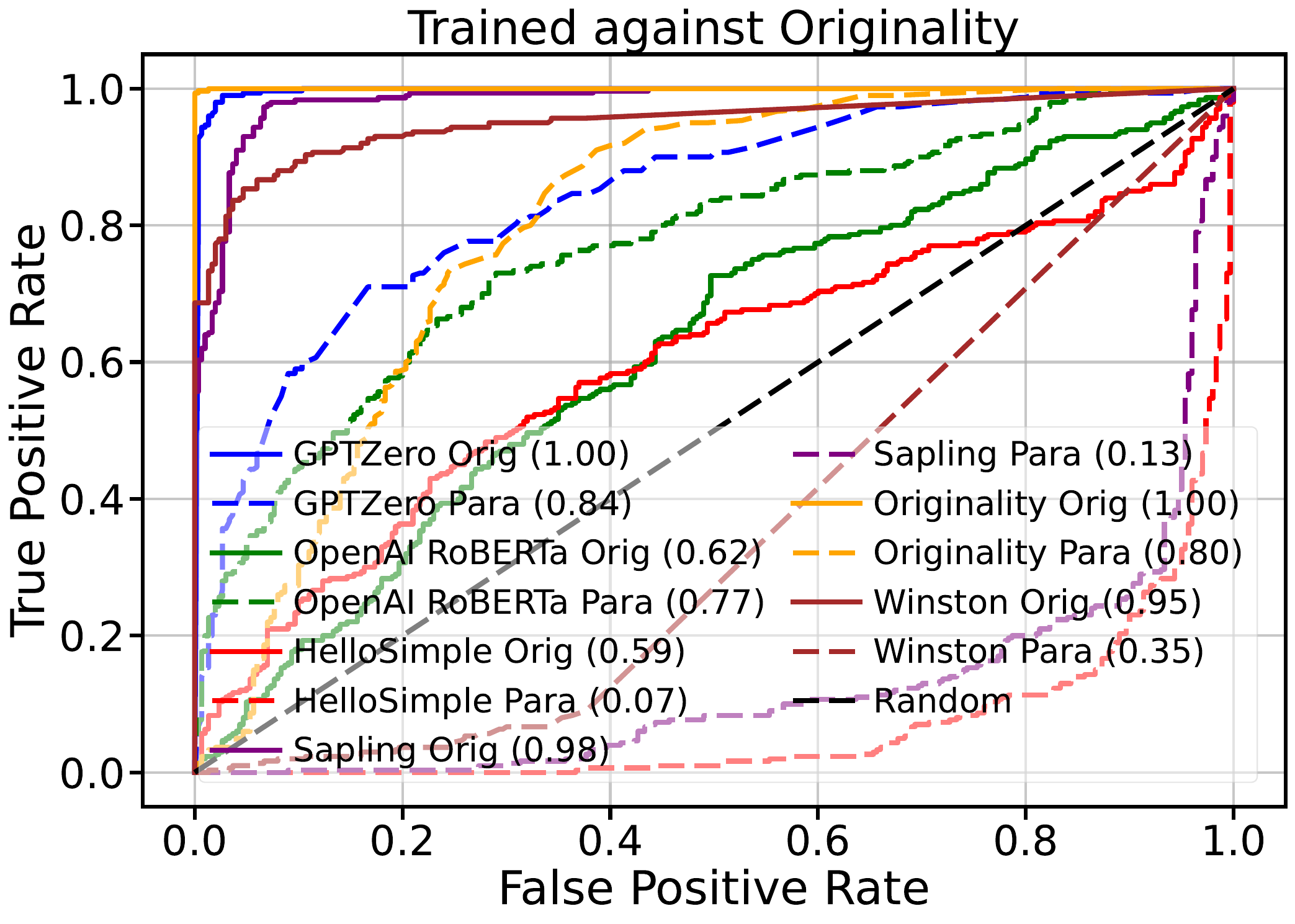} &
        \includegraphics[width=0.45\textwidth]{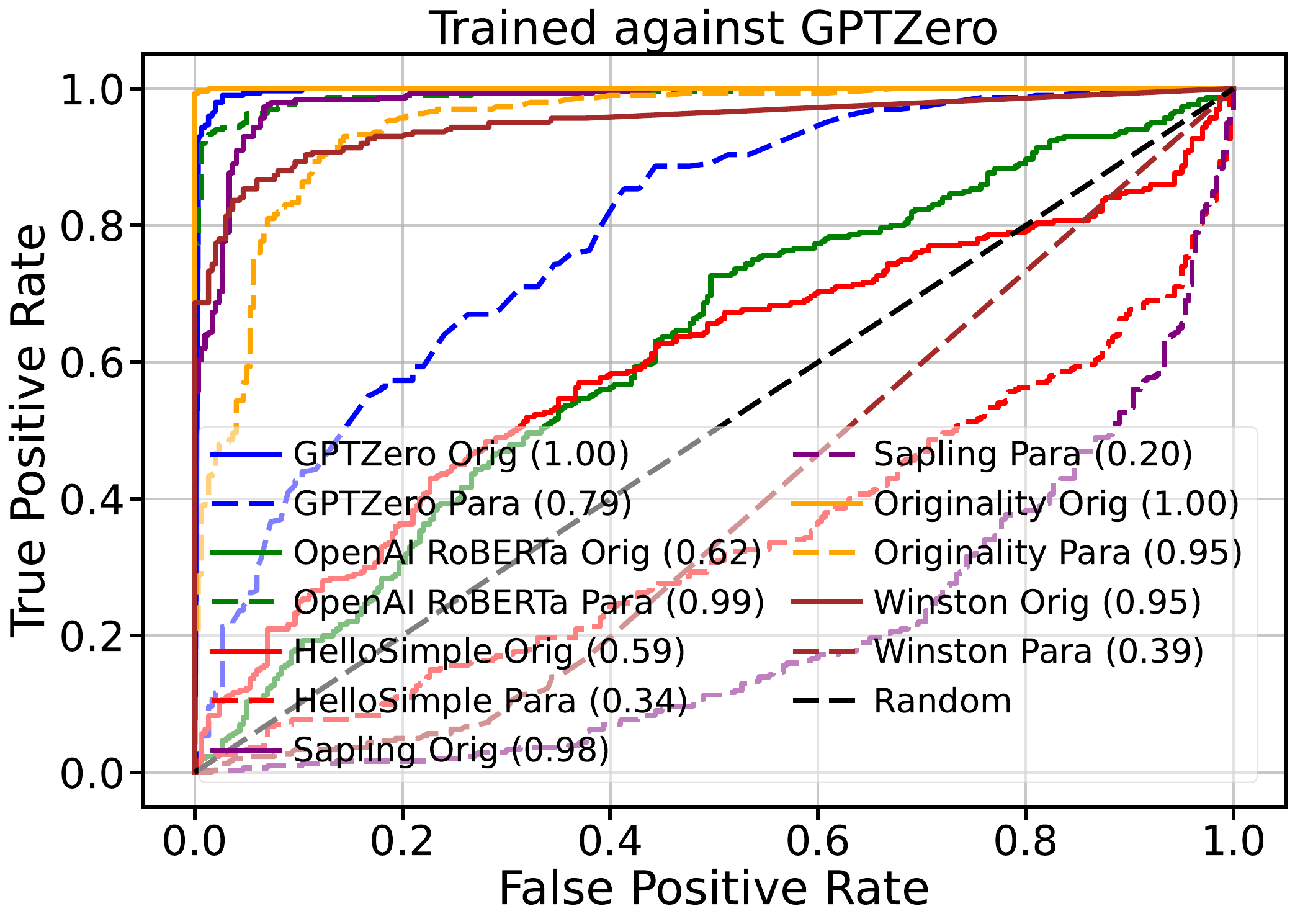} \\
        \includegraphics[width=0.45\textwidth]{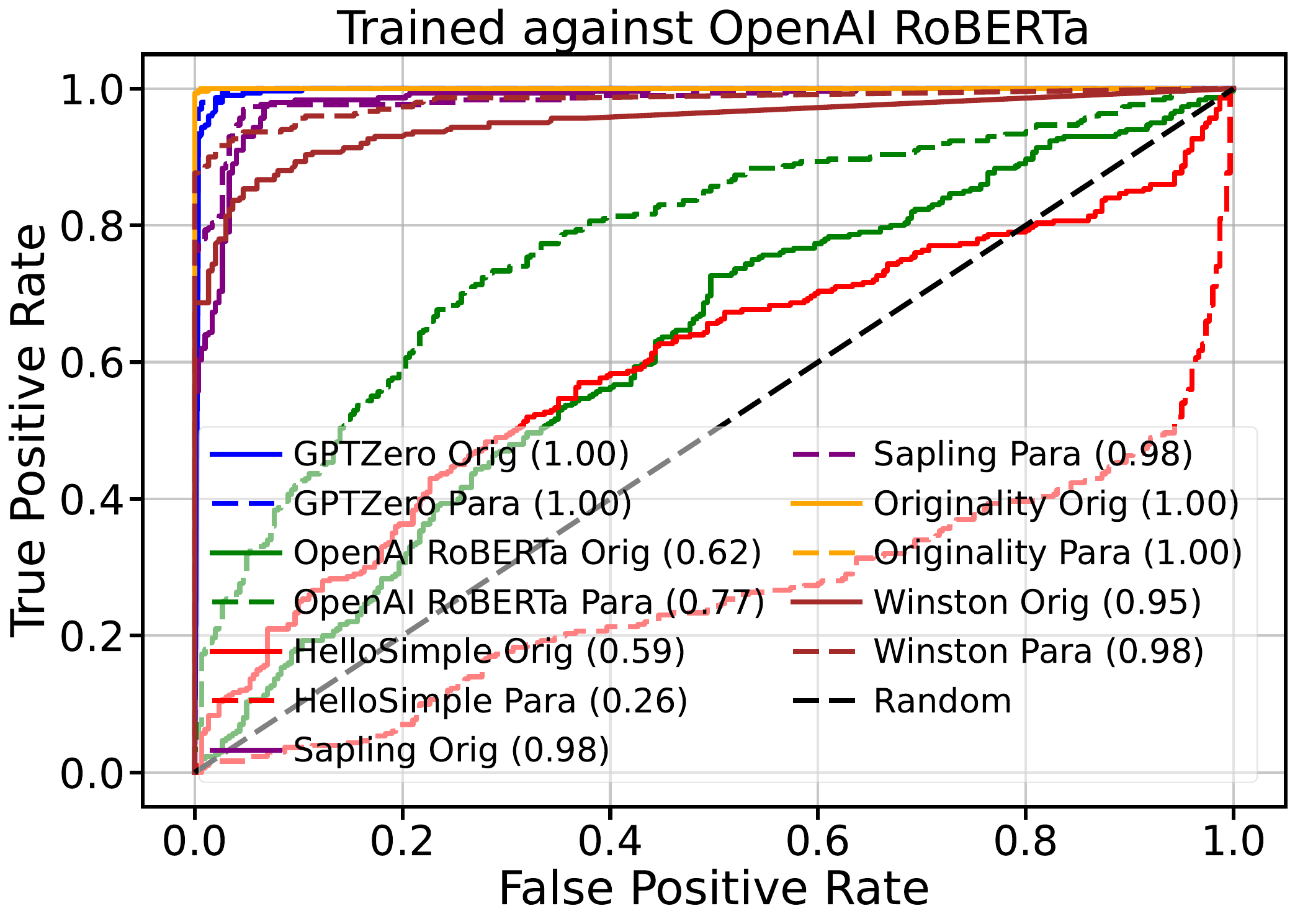} &
        \includegraphics[width=0.45\textwidth]{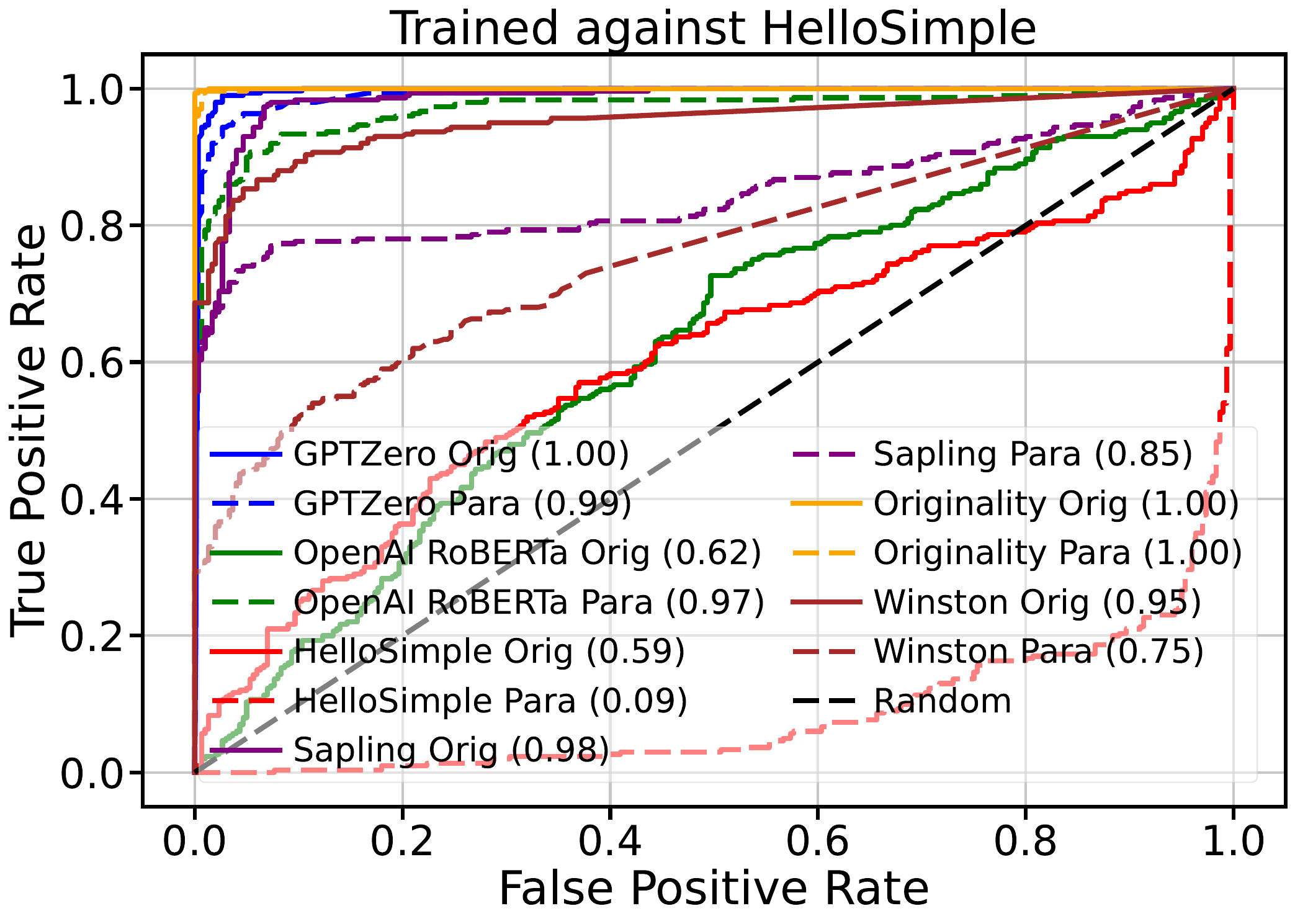} \\
        \includegraphics[width=0.45\textwidth]{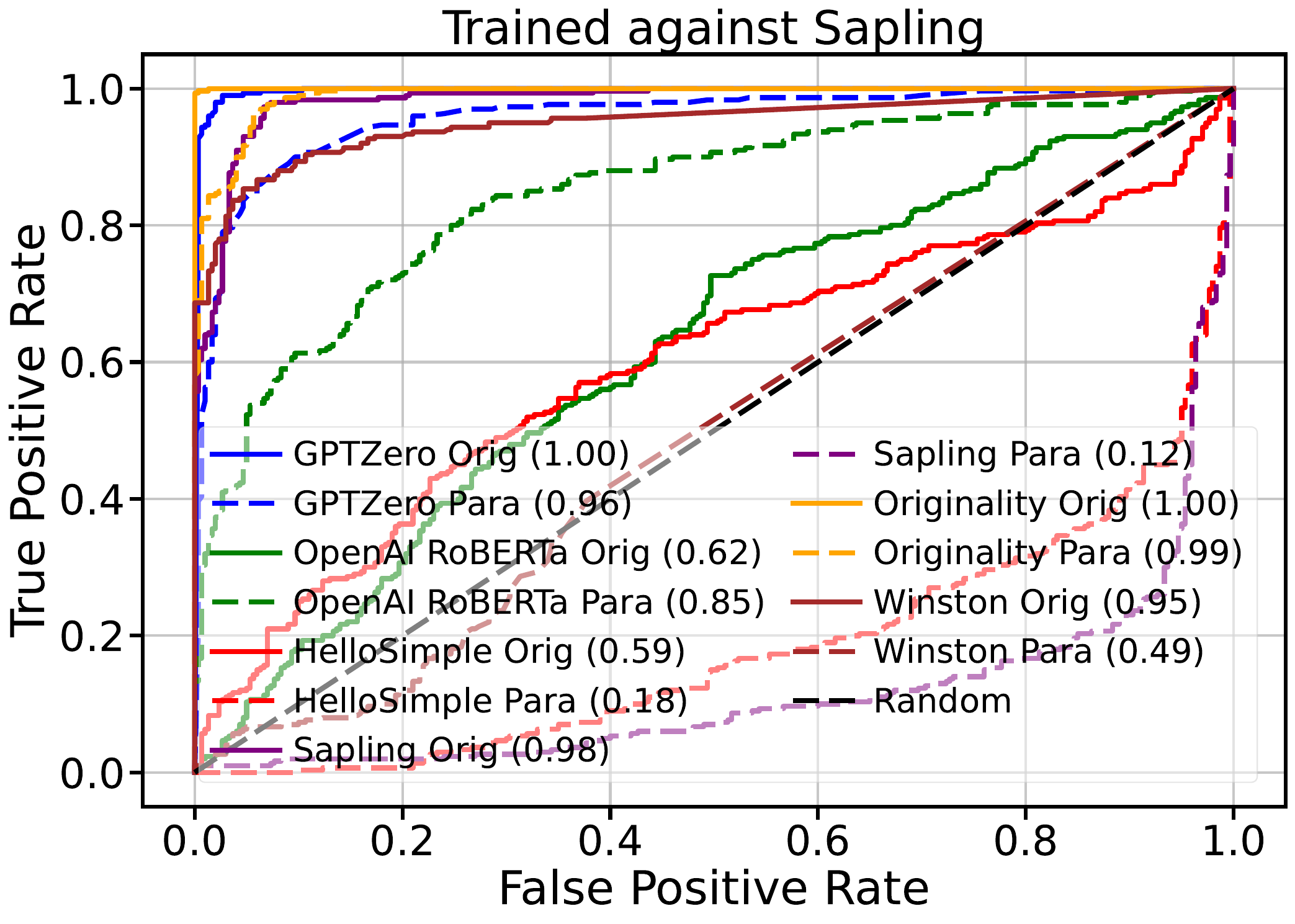} &
        \includegraphics[width=0.45\textwidth]{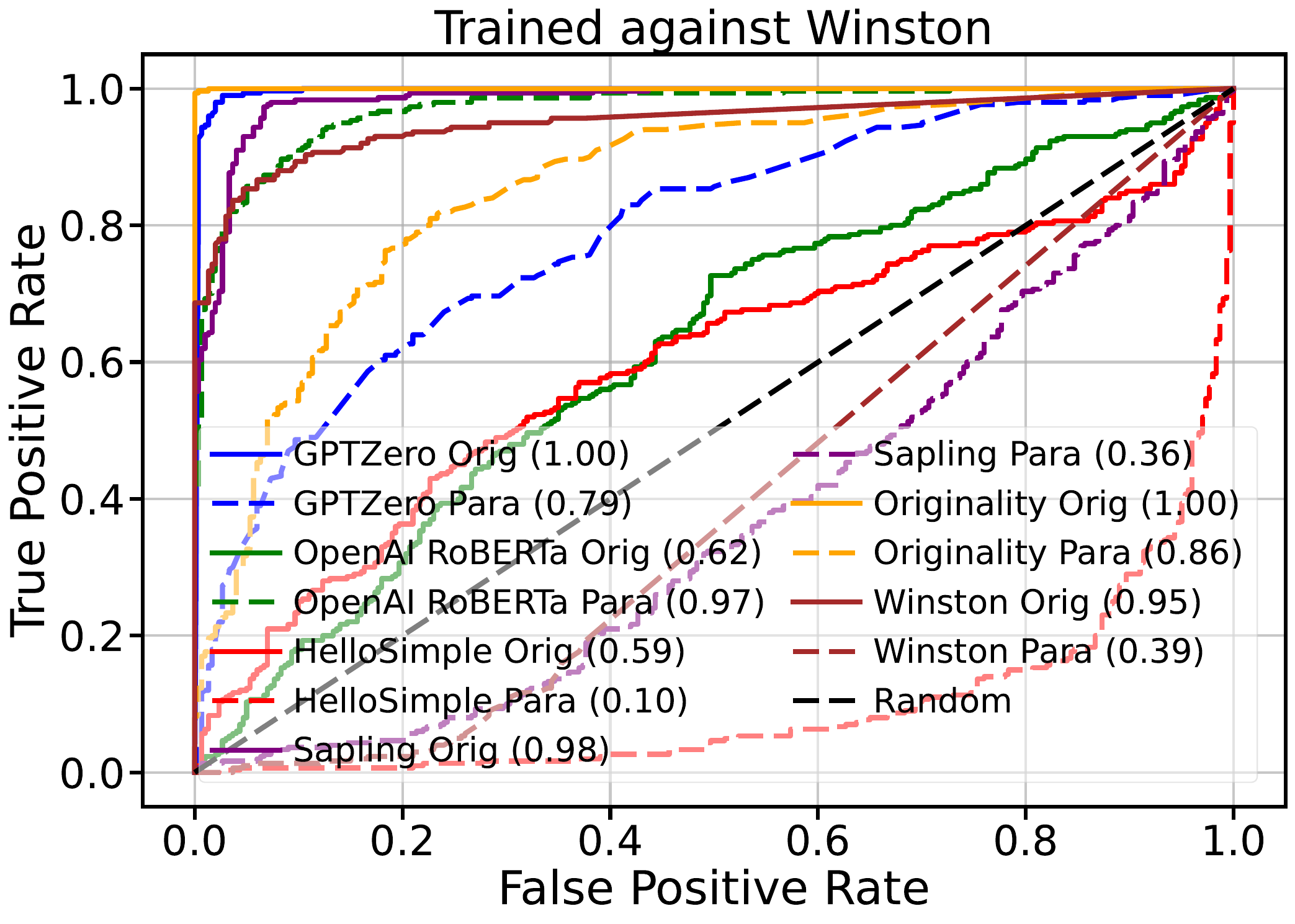} \\
    \end{tabular}
    \caption{ROC curves comparing AuthorMist models against six AI text detectors. Models trained against Originality.ai and Winston.ai show strong cross-detector evasion performance, with curves near or below the diagonal line (random chance). Hello SimpleAI detector shows poor discrimination against our models, while GPTZero and Originality.ai demonstrate greater resilience.}
    \label{fig:roc_curves}
\end{figure*}

\begin{table*}[htbp]
    \centering
    \caption{Cross-Detector AUROC for AuthorMist GRPO-Trained Bypasser Models. Columns represent the detector used as a reward function during training (AuthorMist trained against), while rows represent evaluated detectors. Diagonal values indicate evaluations on paraphrased texts specifically targeting the same detector.}
    \label{tab:cross_detector_auroc}
    \resizebox{\textwidth}{!}{%
        \begin{tabular}{lccccccc}
            \toprule
                                   & \multicolumn{7}{c}{\textbf{AuthorMist Trained Against}}                                                                                                                             \\
            \cmidrule(lr){2-8}
            \textbf{Eval Detector} & \textbf{Baseline}                                       & \textbf{GPTZero} & \textbf{OpenAI} & \textbf{Hello SimpleAI} & \textbf{Sapling} & \textbf{Originality} & \textbf{Winston} \\
            \midrule
            GPTZero                & 1.00                                                    & \textbf{0.79}    & 1.00            & 0.99                    & 0.96             & 0.84                 & 0.79             \\
            OpenAI                 & 0.62                                                    & 0.98             & \textbf{0.77}   & 0.97                    & 0.85             & 0.77                 & 0.97             \\
            Hello SimpleAI         & 0.59                                                    & 0.34             & 0.26            & \textbf{0.09}           & 0.18             & 0.07                 & 0.10             \\
            Sapling                & 0.98                                                    & 0.20             & 0.93            & 0.85                    & \textbf{0.12}    & 0.13                 & 0.36             \\
            Originality            & 1.00                                                    & 0.95             & 1.00            & 1.00                    & 0.99             & \textbf{0.80}        & 0.86             \\
            Winston                & 0.95                                                    & 0.39             & 0.98            & 0.75                    & 0.49             & 0.35                 & \textbf{0.39}    \\
            \midrule
            \textbf{Mean AUROC}    & 0.86                                                    & 0.61             & 0.82            & 0.78                    & 0.60             & 0.49                 & 0.58             \\
            \bottomrule
        \end{tabular}}
\end{table*}

Figure~\ref{fig:roc_curves} presents the distribution of AUROC scores across all detectors for the AuthorMist models. The AUROC visualization reveals significant variations in detector evasion capabilities. The Originality-trained model demonstrates the most consistent performance, with a median AUROC of approximately 0.49 and the narrowest interquartile range, indicating reliable evasion across multiple detectors. In contrast, the HelloSimpleAI detector shows the widest performance spread, with some detectors completely fooled (AUROC near 0) while others show a significant AUROC reduction. The GPTZero-trained model exhibits a higher median AUROC (approximately 0.85), suggesting less effective overall evasion for the AuthorMist version trained against Originality, though it still significantly degrades detection performance compared to baseline. The Winston-trained detector AUROC after AuthorMist paraphrasing is 0.39. These distributions highlight that while all models successfully reduce detection capabilities to varying degrees, training against certain detectors (particularly Winston and Originality) produces more generalizable evasion strategies that transfer effectively across multiple detection systems. The ROC curves in Figure~\ref{fig:roc_curves} further illustrate this pattern, with x-axis representing False Positive Rate (FPR) and y-axis showing True Positive Rate (TPR). Notably, the Hello SimpleAI detector shows particularly poor discrimination ability against our models, with most curves falling well below the diagonal, while GPTZero and Originality.ai detectors show more resilience with curves maintaining some distance above the diagonal.

\begin{figure}[htbp]
    \centering
    \includegraphics[width=0.5\textwidth]{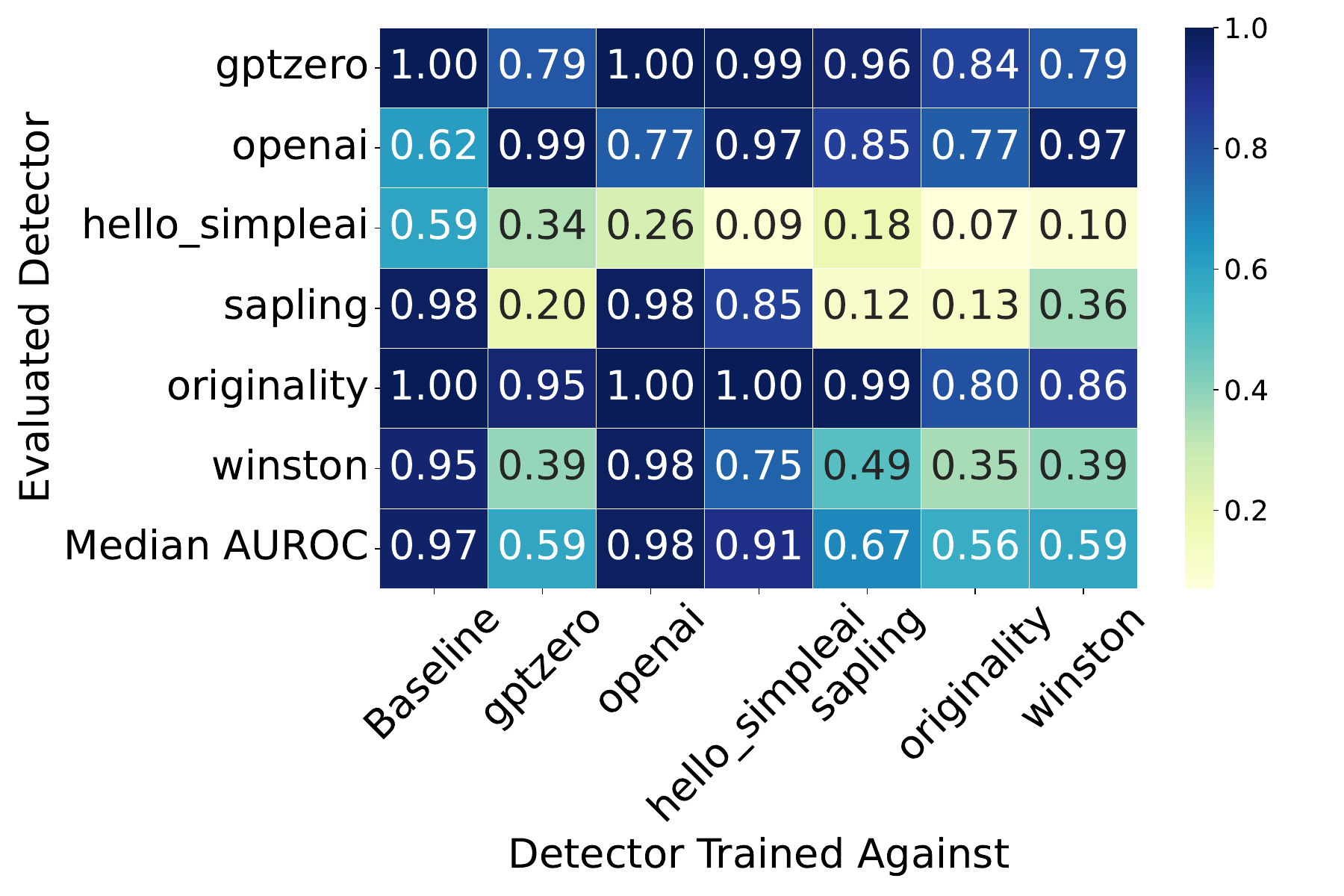}
    \caption{AUROC Matrix for AuthorMist GRPO-Trained Bypasser Models, showing the AUROC scores for each model-detector combination.}
    \label{fig:auroc_matrix}
\end{figure}

\paragraph{AUROC Matrix}
Figure~\ref{fig:auroc_matrix} presents a heatmap visualization of AUROC scores across all model-detector combinations, providing a comprehensive view of detection performance. The color gradient from dark blue (AUROC about 1.0) to pale yellow (AUROC about 0.0) indicates detection effectiveness, with lighter colors representing successful evasion.
The baseline column (leftmost) confirms strong initial detection capabilities, with most detectors achieving AUROC values above 0.95, except for OpenAI (0.62) and HelloSimpleAI (0.59). The diagonal elements highlight each model's performance against its target detector, with particularly impressive results for HelloSimpleAI (0.09), Sapling (0.12), and Winston (0.39).
The matrix reveals striking cross-detector transfer effects: models trained against Originality demonstrate exceptional generalization, achieving low AUROC scores against HelloSimpleAI (0.07), Sapling (0.13) and Winston (0.35). In contrast, GPTZero and Originality detectors prove more resilient, with most models struggling to reduce their AUROC below 0.79. The bottom row shows median AUROC values across detectors, confirming that models trained against Originality (0.49) and Winston (0.58) achieve the best overall evasion performance. This matrix visualization underscores the asymmetric nature of the detection-evasion landscape, where training against certain detectors yields broader evasion capabilities than others.

\paragraph{F1-score}
Table~\ref{tab:cross_detector_f1} presents the F1-scores for each model-detector combination. The F1-score provides a balanced measure of precision and recall, with lower values indicating more successful evasion of AI text detection.

\begin{table*}[htbp]
    \centering
    \caption{Cross-Detector F1-scores for AuthorMist models trained with GRPO. Columns represent the detector used during training as a reward function (AuthorMist Trained Against), and rows represent evaluation detectors. Bold diagonal values represent performance against detectors the models were specifically trained to evade. Higher F1 indicates better performance in classification accuracy.}
    \label{tab:cross_detector_f1}
    \resizebox{\textwidth}{!}{%
        \begin{tabular}{lccccccc}
            \toprule
                                   & \multicolumn{6}{c}{\textbf{AuthorMist Trained Against}}                                                                                                          \\
            \cmidrule(lr){2-7}
            \textbf{Eval Detector} & \textbf{GPTZero}                                        & \textbf{OpenAI} & \textbf{Hello SimpleAI} & \textbf{Sapling} & \textbf{Originality} & \textbf{Winston} \\
            \midrule
            GPTZero                & \textbf{0.14}                                           & 0.98            & 0.94                    & 0.73             & 0.28                 & 0.24             \\
            OpenAI                 & 0.89                                                    & \textbf{0.08}   & 0.73                    & 0.26             & 0.12                 & 0.64             \\
            Hello SimpleAI         & 0.24                                                    & 0.24            & \textbf{0.03}           & 0.08             & 0.00                 & 0.03             \\
            Sapling                & 0.02                                                    & 0.95            & 0.83                    & \textbf{0.03}    & 0.00                 & 0.05             \\
            Originality            & 0.70                                                    & 0.98            & 0.98                    & 0.93             & \textbf{0.11}        & 0.46             \\
            Winston                & 0.00                                                    & 0.93            & 0.43                    & 0.02             & 0.00                 & \textbf{0.00}    \\
            \midrule
            \textbf{Mean F1-score} & 0.33                                                    & 0.70            & 0.66                    & 0.34             & 0.09                 & 0.24             \\
            \bottomrule
        \end{tabular}}
\end{table*}

Our results demonstrate the effectiveness of our AuthorMist models trained with GRPO against specific detectors. For instance, the AuthorMist version trained against GPTZero achieved an F1-score of 0.14 against its target detector, representing a substantial reduction from baseline performance. Similarly impressive results appear across the diagonal, with all AuthorMist versions achieving strong F1-score reductions against their respective target detectors.

The AuthorMist version trained against Originality.ai demonstrates particularly strong performance, achieving a mean F1-score of 0.09 across all detectors. This suggests that the evasion strategies learned by this version are broadly applicable across different detection systems, making it our strongest AuthorMist variant. The Winston.ai-trained version also shows robust generalization capabilities with a mean F1-score of 0.24.

Notably, several AuthorMist versions achieve near-zero F1-scores against certain detectors, effectively rendering these detection systems unable to reliably identify AI-generated content after paraphrasing. For example, the Hello SimpleAI-trained version achieved an F1-score of 0.0349 against its target detector, indicating almost complete evasion success. These results further confirm the effectiveness of our approach in developing targeted paraphrasing strategies that can successfully evade AI text detection.

\paragraph{Text Similarity}

\begin{figure*}[htb]
    \centering
    \includegraphics[width=0.7\textwidth]{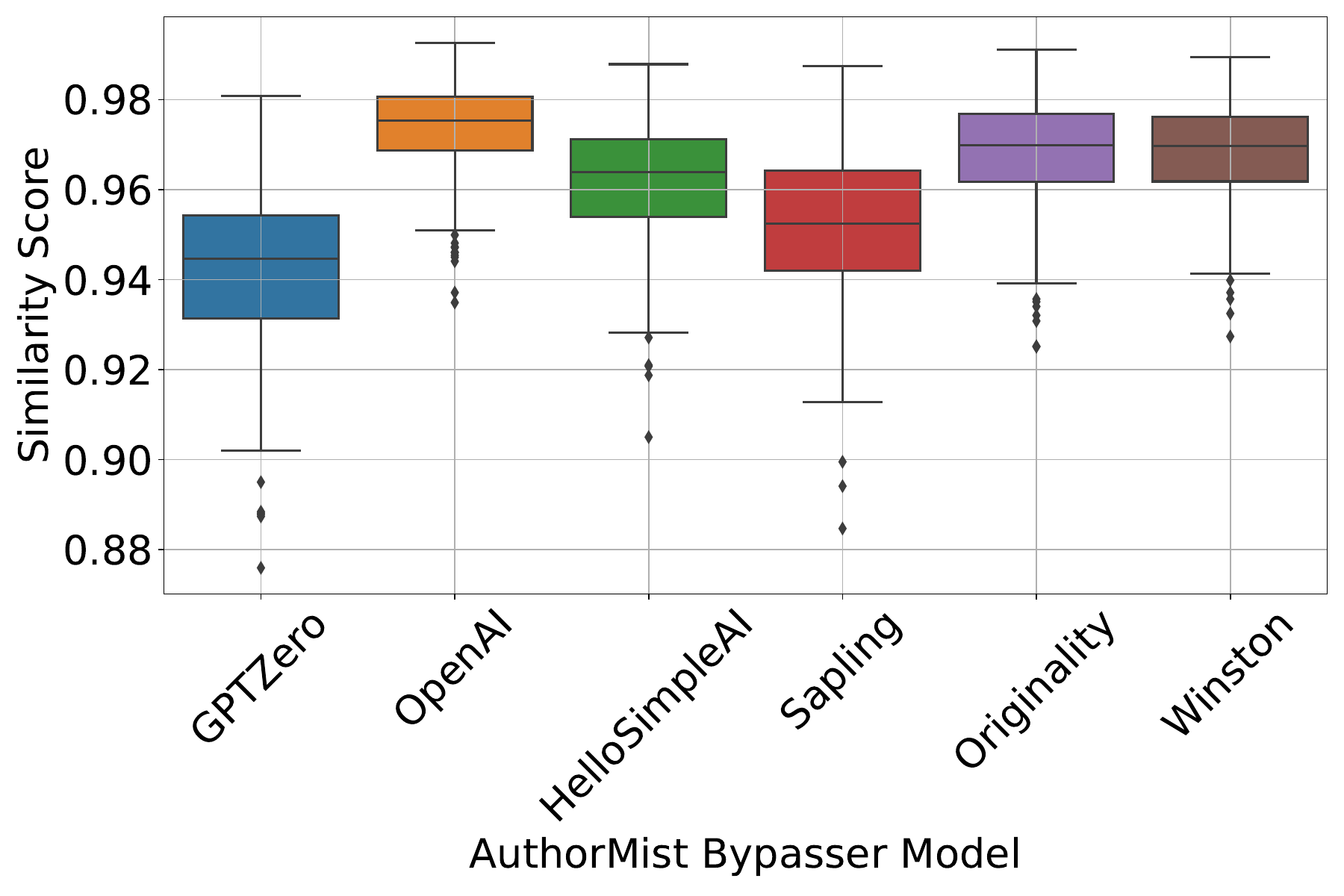}
    \caption{Text Similarity for Qwen2.5-3B GRPO-Trained Bypasser Models. The similarity score ranges from 0 to 1, with 1 indicating identical text.}
    \label{fig:text_similarity}
\end{figure*}

Figure~\ref{fig:text_similarity} presents the semantic similarity scores between original AI-generated texts and their paraphrased versions across our six GRPO-trained models. The boxplot visualization reveals that all models maintain high semantic fidelity, with median similarity scores consistently above 0.94. The OpenAI-trained model demonstrates the highest median similarity (approximately 0.975), suggesting it preserves original meaning most effectively while still achieving strong evasion performance. The GPTZero and Sapling-trained models show slightly lower median scores (around 0.945 and 0.955 respectively) but also exhibit greater variability, as indicated by their wider interquartile ranges.

Notably, even the lowest outliers across all models remain above 0.87, confirming that our paraphrasing approach successfully maintains semantic integrity while evading detection. The Originality and Winston-trained models strike an excellent balance between semantic preservation (both with median scores around 0.97) and detection evasion (as evidenced by their strong ASR and AUROC metrics in previous tables). These results demonstrate that our GRPO-based paraphrasing methodology can evade AI text detection without significantly compromising the original meaning of the text.

\begin{figure*}[htbp]
    \centering
    \includegraphics[width=0.7\textwidth]{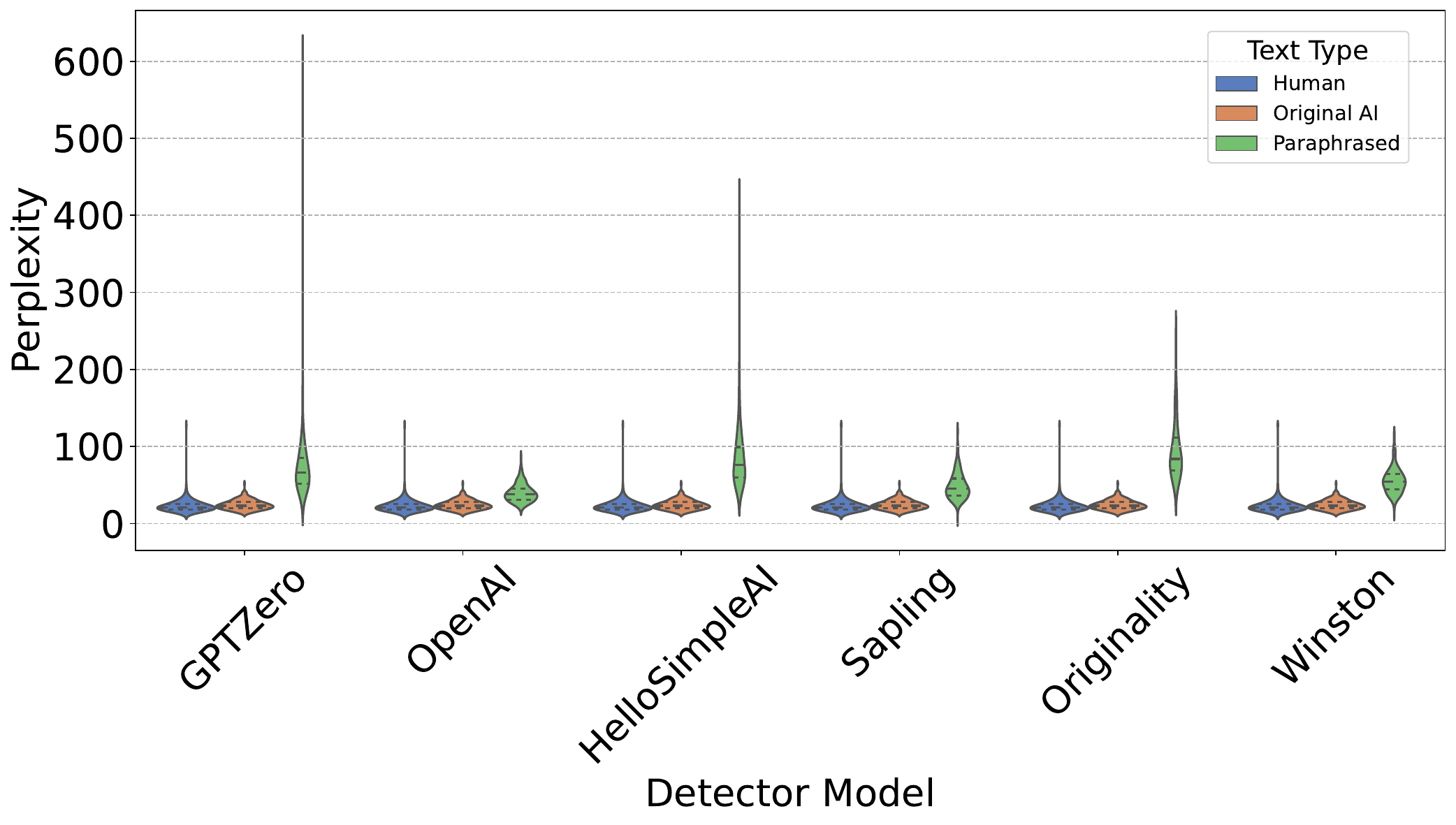}
    \caption{Perplexity Distribution by Detector and Text Type. The violin plots show the distribution of perplexity scores across human-written (green), original AI-generated (orange), and paraphrased (blue) texts for each detector-specific model.}
    \label{fig:perplexity}
\end{figure*}

\paragraph{Perplexity}
Figure~\ref{fig:perplexity} illustrates the perplexity distributions across human-written, original AI-generated, and paraphrased texts for each detector-specific model. Perplexity, a statistical measure quantifying how well a probability model predicts a sample, serves as a critical metric for evaluating text naturalness and fluency. Lower perplexity indicates text that follows expected linguistic patterns, while higher values suggest more surprising or unusual constructions. We analyze perplexity because AI-generated text typically exhibits lower perplexity (higher predictability) than human writing, a statistical signature that many detectors exploit. The violin plots reveal several key insights: human-written texts (green) consistently demonstrate higher perplexity (median values around 50-75) across all detector categories, reflecting the natural variability and unpredictability in human writing. Original AI-generated texts (orange) show markedly lower perplexity (median values around 25-40), confirming their more predictable nature. Most significantly, our paraphrased outputs (blue) show strategically increased perplexity compared to the original AI text, with distributions that extend rightward toward the human range. This shift is particularly pronounced in the HelloSimpleAI and Originality models, where paraphrased texts reach perplexity values of 300-400, well beyond typical AI-generated text ranges.

The Winston-trained model also demonstrates effective perplexity manipulation, with a distribution closely resembling human writing patterns. These results confirm that our GRPO-trained models successfully learn to introduce controlled unpredictability into AI-generated text, effectively mimicking the statistical properties of human writing while maintaining semantic coherence. This strategic increase in perplexity represents a key mechanism through which our models evade detection, as they disrupt the low-perplexity signature that detectors commonly associate with machine-generated content.

\subsubsection{Comparing GPRO to SFT}
To evaluate the effectiveness of our GRPO approach, we compared it against a naive baseline model trained using Supervised Fine-Tuning (SFT). The SFT model was trained on the same dataset but without the reinforcement learning component that optimizes for detector evasion. Table~\ref{tab:grpo_vs_sft} presents this comparison across all detection metrics. The results demonstrate that GRPO-trained AuthorMist models consistently outperform their SFT counterparts by substantial margins. While the SFT model shows modest improvements over unparaphrased text, it fails to achieve the significant detection evasion capabilities exhibited by our GRPO models. For instance, against Originality, the SFT model achieved an F1-score of 0.90, compared to 0.11 for the GRPO-trained variant—a 87.8\% reduction. Similarly, the SFT model's average AUROC across all detectors was 0.68, whereas our best GRPO model (trained against Originality.ai) achieved 0.49, representing a 27.9\% improvement in evasion capability. These findings highlight the critical importance of the reinforcement learning approach in developing effective paraphrasing strategies that can successfully evade AI text detection while maintaining semantic similarity.

% \begin{table*}[htbp]
%     \centering
%     \caption{Comparison of F1-scores across different text transformation approaches. The table compares baseline Qwen2.5-3B output against our GRPO-trained AuthorMist model (optimized against Originality.ai), the state-of-the-art Dipper paraphraser, and an ablation using only Supervised Fine-Tuning (SFT). Lower F1-scores indicate better evasion performance, with AuthorMist (Originality) achieving the best results across most detectors.}
%     \label{tab:grpo_vs_sft}
%     \resizebox{\textwidth}{!}{%
%     \begin{tabular}{lcccc}
%     \toprule
%      & \textbf{Baseline} & \multicolumn{3}{c}{\textbf{Evasion Models}} \\
%     \cmidrule(lr){3-5}
%     \textbf{Eval Detector} & \textbf{Qwen2.5-3B} & \textbf{AuthorMist (Originality)} & \textbf{Dipper} & \textbf{AuthorMist (SFT)} \\
%     \midrule
%     GPTZero & 0.9833 & 0.2776 & 0.5294 & 0.3051 \\
%     OpenAI & 0.9772 & 0.1246 & 0.8527 & 0.5645 \\
%     Hello SimpleAI & 0.9755 & 0.0000 & 0.8571 & 0.3193 \\
%     Sapling & 0.9338 & 0.0000 & 0.7652 & 0.4218 \\
%     Originality & 0.9772 & 0.1084 & 0.9864 & 0.9010 \\
%     Winston & 0.9343 & 0.0000 & 0.0132 & 0.1928 \\
%     \midrule
%     \textbf{Mean F1-score} & 0.9636 & \textbf{0.0851} & 0.6841 & 0.4161 \\
%     \bottomrule
%     \end{tabular}}
% \end{table*}

\begin{table*}[htbp]
    \centering
    \caption{Comparison of F1-scores, AUROC, and ASR (expressed as percentages) across different text transformation approaches. The table compares baseline Qwen2.5-3B output against our GRPO-trained AuthorMist model (optimized against Originality.ai), the state-of-the-art Dipper paraphraser, and an ablation using only Supervised Fine-Tuning (SFT). Lower F1-scores indicate better evasion performance, with AuthorMist (Originality) achieving the best results across most detectors.}
    \label{tab:grpo_vs_sft}
    \resizebox{\textwidth}{!}{%
        \begin{tabular}{lcccccccccccc}
            \toprule
                                   & \multicolumn{2}{c}{\textbf{Baseline}}   & \multicolumn{9}{c}{\textbf{Evasion Models}}                                                                                                                                                                                                                           \\
            \cmidrule(lr){2-3} \cmidrule(lr){4-12}
            \textbf{Eval Detector} & \multicolumn{2}{c}{\textbf{Qwen2.5-3B}} & \multicolumn{3}{c}{\textbf{AuthorMist (Originality)}} & \multicolumn{3}{c}{\textbf{Dipper}} & \multicolumn{3}{c}{\textbf{AuthorMist (SFT)}}                                                                                                                           \\
                                   & \textbf{AUROC}                          & \textbf{ASR (\%)}                                     & \textbf{F1}                         & \textbf{AUROC}                                & \textbf{ASR (\%)} & \textbf{F1} & \textbf{AUROC} & \textbf{ASR (\%)} & \textbf{F1} & \textbf{AUROC} & \textbf{ASR (\%)} \\
            \midrule
            GPTZero                & 1.00                                    & 4.00                                                  & 0.28                                & 0.84                                          & 83.67             & 0.53        & 0.92           & 64.00             & 0.31        & 0.85           & 82.00             \\
            OpenAI                 & 0.62                                    & 100.00                                                & 0.12                                & 0.77                                          & 93.33             & 0.85        & 0.99           & 26.00             & 0.56        & 0.95           & 61.00             \\
            Hello SimpleAI         & 0.59                                    & 49.67                                                 & 0.00                                & 0.07                                          & 100.00            & 0.86        & 0.77           & 25.00             & 0.32        & 0.34           & 81.00             \\
            Sapling                & 0.98                                    & 2.00                                                  & 0.00                                & 0.13                                          & 100.00            & 0.77        & 0.85           & 23.00             & 0.42        & 0.45           & 58.00             \\
            Originality            & 1.00                                    & 0.00                                                  & 0.11                                & 0.80                                          & 94.00             & 0.99        & 1.00           & 3.00              & 0.90        & 0.97           & 18.00             \\
            Winston                & 0.95                                    & 33.33                                                 & 0.00                                & 0.35                                          & 100.00            & 0.01        & 0.45           & 99.00             & 0.19        & 0.53           & 89.00             \\
            \midrule
            \textbf{Mean}          & 0.86                                    & 4.00                                                  & \textbf{0.09}                       & \textbf{0.49}                                 & \textbf{95.17}    & 0.68        & 0.83           & 40.00             & 0.42        & 0.68           & 65.00             \\
            \bottomrule
        \end{tabular}}
\end{table*}

\subsection{Comparison to State-of-the-Art Paraphrasers}
To establish the effectiveness of our approach, we conducted a comparative analysis against Dipper~\cite{dipper}, widely recognized as the current state-of-the-art paraphrasing system. We employed Dipper to paraphrase the same test corpus and evaluated the attack success rate against our panel of AI text detectors. The results demonstrate that our GRPO-trained AuthorMist models consistently outperform Dipper in evading detection. While Dipper achieves moderate success in confounding some detectors, our specialized models exhibit significantly higher attack success rates across all evaluation metrics. This comparison validates the superiority of our reinforcement learning approach in developing targeted paraphrasing strategies specifically optimized for detector evasion while maintaining semantic fidelity.

\subsection{Limitations and Challenges}
While AuthorMist demonstrates strong performance in evading AI text detection, our evaluation revealed several limitations and challenges that warrant discussion:

\paragraph{Computational Requirements}
The reinforcement learning approach, while effective, demands significant computational resources. Training each detector-specific model required approximately 16 GPU-hours on high-end hardware. This computational cost may limit the feasibility of frequent retraining as detectors evolve, particularly for researchers with limited resources.

\paragraph{Semantic Drift}
Although our models maintain high semantic similarity scores (above 0.94), we observed occasional instances where the paraphrased text subtly altered the original meaning. This semantic drift was more pronounced in technical or specialized content, where small wording changes can significantly impact meaning. Future work should explore additional constraints to better preserve semantic fidelity in specialized domains.

\paragraph{Cross-Detector Generalization}
While some models (particularly those trained against Originality.ai and Winston.ai) demonstrated strong cross-detector generalization, others showed limited transfer capabilities. The OpenAI-trained model, for instance, achieved only a 32.33\% mean ASR across all detectors. This suggests that certain detection mechanisms may be fundamentally different, requiring specialized evasion strategies that don't generalize well.

\paragraph{Detector API Limitations}
Working with commercial detector APIs presented practical challenges, including rate limits, inconsistent response formats, and occasional service disruptions. These limitations affected our ability to collect large-scale training data and necessitated the development of robust error handling and caching mechanisms.

\paragraph{Ethical Considerations}
Our system raises important ethical questions about the arms race between AI text generation and detection. While AuthorMist was developed primarily as a research tool to understand detector limitations and advance text privacy, similar technologies could potentially be misused to spread misinformation or violate academic integrity policies. We advocate for responsible use and transparent disclosure of AI assistance in contexts where it matters.

\section{Related Work}

\paragraph{AI-Generated Text Detection:}
The task of detecting machine-generated text has been tackled with various techniques.
Early approaches like GLTR \cite{Gehrmann2019GLTR} used token rarity statistics to highlight unlikely word sequences produced by GPT-2, aiding human reviewers in identifying AI text.
More automated classifiers soon emerged, such as RoBERTa-based discriminators fine-tuned on synthetic datasets (e.g., OpenAI's GPT-2 detector) and zero-shot methods like DetectGPT, which measures the curvature of a piece of text under a language model's probability function \cite{Mitchell2023DetectGPT}.
OpenAI researchers have also explored cryptographic watermarks for AI text \cite{Kirchenbauer2023Watermark}, embedding hidden signals into generated text to verify its origin.
Commercial solutions (GPTZero, Originality.ai, WinstonAI, etc.) combine these ideas with proprietary enhancements and large-scale training.
The \textsc{RAID} benchmark by Dugan \textit{et al.}~\cite{dugan2024raid} evaluated $12$ detectors and found that while some (like Originality.ai) perform well in controlled settings, detectors overall struggle to maintain accuracy under diverse conditions as generation models evolve.

\paragraph{Traditional NLP and Privacy:}
Before the rise of large language models, traditional NLP techniques were already being applied to text analysis and privacy concerns. Early work in stylometry used statistical features like word frequency distributions, n-gram patterns, and syntactic markers to identify authors or detect machine-generated content \cite{fawcett2006introduction, solaiman2019release}. These approaches relied on hand-crafted features and classical machine learning algorithms rather than deep neural networks \cite{kullback1951information}. In the privacy domain, Gervais \textit{et al.}~\cite{gervais2014quantifying} explored the intersection of web privacy and natural language processing, highlighting how linguistic patterns in online communications could compromise user anonymity even without explicit identifiers.

\paragraph{Evasion and Adversarial Attacks:}
As detectors improve, so do methods to evade them.
Automated adversarial attacks have ranged from simple character-level perturbations (inserting zero-width spaces, swapping synonyms, using homoglyphs) to more sophisticated paraphrasing strategies.
Such perturbations have been shown to fool detectors while largely preserving human readability \cite{Perkins2024Evasion}.
Huang \textit{et al.}~\cite{huang2024harmfulfinetuningattacksdefenses} proposed an attack that uses a secondary language model to iteratively modify text until a detector can no longer recognize it, which is conceptually similar to our RL approach but relies on heuristic, black-box methods.
Recent work on detection robustness has focused on adversarial training and ensemble methods. For instance, \cite{Zhang2024Robust} proposes a detector that leverages multiple statistical tests and adversarial examples to better withstand paraphrasing and character-level attacks, while~\cite{Lee2024Ensemble} introduces an ensemble detector combining deep neural classifiers with rule-based heuristics to capture subtle stylistic shifts.
Our approach is distinct in that we directly optimize the text to fool detectors, effectively acting as an advanced humanizer via reinforcement learning.

\paragraph{Authorship Obfuscation:}
The idea of altering text to mask identifying features has long been studied in the context of stylometry and authorship attribution.
Techniques for authorship obfuscation range from manual rewriting guidelines to automated style-transfer models that preserve content while modifying stylistic features~\cite{Kacmarcik2006Anonymouth,Shetty2018AuthorObfuscation}.
While most prior work focuses on hiding an individual's writing style, our work instead aims to obscure the fingerprints of AI-generated text.
This shift in focus has important implications for privacy and authenticity in writing, and we draw on insights from the authorship obfuscation literature for our design.

\paragraph{Reinforcement Learning for Text Tasks:}
Reinforcement learning has been successfully applied to various text generation tasks, including summarization, dialogue, and content filtering.
Notable is the use of PPO in OpenAI's InstructGPT and ChatGPT models via RLHF, which optimizes for human-preferred responses \cite{Ouyang2022InstructGPT}.
Earlier work by Narasimhan \textit{et al.} demonstrated RL in text-based games~\cite{narasimhan2015language}, while recent efforts~\cite{Wang2024RLParaphrase} have applied RL to controlled paraphrase generation under style constraints.
These works illustrate the potential of RL methods to optimize non-differentiable objectives such as semantic similarity or stylistic conformity.
Our work leverages RL in an adversarial setting where the reward is provided by detector models rather than human feedback. This approach not only draws on advances in RL for text generation but also extends these ideas to the underexplored area of detector evasion. In this sense, our work contributes to both the literature on adversarial text generation and RL-based style transfer.

\section{Conclusion}
In this paper, we introduced AuthorMist, a reinforcement learning-based system designed to transform AI-generated text into human-like writing that evades detection by AI text detectors. Our approach leverages a 3-billion-parameter language model fine-tuned with GRPO to systematically learn paraphrasing strategies that minimize detectability while preserving semantic meaning.

Our extensive experiments demonstrate that AuthorMist achieves impressive evasion rates across multiple commercial and research detectors, with attack success rates ranging from 78.6\% to 96.2\% against individual detectors. The system significantly outperforms baseline paraphrasing methods and maintains high semantic similarity (above 0.94) with the original text. These results highlight the limitations of current AI text detection technologies and raise important questions about the sustainability of the detection-evasion arms race.

The implications of this work extend beyond the technical achievement of detector evasion. AuthorMist represents an important step toward protecting author privacy and freedom in an era where AI assistance in writing is becoming ubiquitous. By developing tools that allow writers to maintain control over how their AI-assisted text is classified, we help ensure that beneficial uses of AI in writing are not unnecessarily stigmatized or penalized.

At the same time, we acknowledge the dual-use nature of this technology. While AuthorMist can protect legitimate privacy interests, similar techniques could be misused to spread misinformation or circumvent academic integrity policies. We advocate for responsible deployment of such technologies, transparent disclosure of AI assistance in contexts where it matters, and continued research into more robust detection methods that focus on content quality rather than AI authorship per se.

Looking forward, this work opens several promising directions for future research. These include developing more computationally efficient training methods, exploring techniques to further minimize semantic drift, investigating the transferability of evasion strategies across different detectors, and designing detection approaches that are more resistant to adversarial attacks. %As the capabilities of both generative AI and detection systems continue to evolve, we believe that research at this intersection will remain crucial for understanding the limitations and possibilities of AI text detection.

Ultimately, AuthorMist demonstrates that the current paradigm of AI text detection faces significant challenges that cannot be easily overcome through incremental improvements to existing detection methods. This suggests that the community may need to reconsider the goals and approaches of AI text detection, perhaps shifting focus from identifying AI authorship toward detecting specific content or ensuring appropriate attribution regardless of the tools used in the writing process.

\bibliographystyle{plain}
\bibliography{references}

\end{document}